\documentclass[12pt,preprint]{aastex}
\begin{document}

\title{21 cm Signals from Early Ionizing Sources}

\author{Jiren Liu\altaffilmark{1}, Jing-Mei Qiu\altaffilmark{2},
        Long-Long Feng\altaffilmark{3,4},
       Chi-Wang Shu\altaffilmark{2},
       and Li-Zhi Fang\altaffilmark{1}}

\altaffiltext{1}{Department of Physics, University of Arizona,
Tucson, AZ 85721}

\altaffiltext{2}{Division of
Applied Mathematics, Brown University, Providence, RI 02912}

\altaffiltext{3}{Purple Mountain Observatory, Nanjing, 210008,
P.R. China.}

\altaffiltext{4}{National Astronomical Observatories, Chinese
Academy of Science, Chao-Yang District, Beijing 100012, P.R.
China}

\begin{abstract}

We investigate the 21 cm signals from the UV
ionizing sources in the reionization epoch. The formation and
evolution of 21 cm emission and absorption regions depend
essentially on the kinetics of photons in the physical and frequency
spaces. To solve the radiative transfer equation, we use the WENO
algorithm, which is effective to capture the sharp ionization
profile and the cut-off at the front of light $(r=ct)$ and to handle
the small fraction of neutral hydrogen and helium in the ionized
sphere. We show that a spherical shell of 21 cm emission and
absorption will develop around a point source once the speed of the
ionization front (I-front) is significantly lower than the speed of
light. The 21 cm shell extends from the I-front to the front of
light; its inner part is the emission region and its outer part is the
absorption region. The 21 cm emission region depends
strongly on the intensity, frequency-spectrum and life-time of the
UV ionizing source. At redshift $1+z=20$, for a UV ionizing source
with an intensity $\dot{E}\simeq10^{45}{\rm ergs}^{-1}$ and a power law
spectrum $\nu^{-\alpha}$ with $\alpha=2$, the emission region
has a comoving size of 1 - 3 Mpc at time $\simeq 2$ Myr. Nevertheless, the
emission regions are very small, and would be erased by thermal
broadening if the intensity is less than $\dot{E}\simeq
10^{43}{\rm ergs}^{-1}$, the frequency spectrum is thermal at
temperature $T\simeq 10^5$ K, or the frequency spectrum is a power law 
with $\alpha\geq 3$. On the other hand, the 21 cm absorption regions
are developed in all these cases. For a source of short life-time,
no 21 cm emission region can be formed if the source dies out before
the I-front speed is significantly lower than the speed of light.
Yet, a 21 cm absorption region can form and develop even after the
emission of the source ceases.

\end{abstract}

\keywords{cosmology: theory - intergalactic medium - radiation transfer -
methods: numerical }

\newpage

\section{Introduction}

The detection of redshifted 21 cm signals from the early universe is
attracting many attentions in the study of cosmology, because it 
provides a window to probe the baryonic gas and the first
generation of light sources in the cosmic dark ages (e.g. Furlanetto et
al. 2006). Observational projects of highly sensitive meter radio
telescopes, such as the Low Frequency Array (LOFAR) (Rottgering 2003), the
Square Kilometer Array (SKA) (van de Weygaert \& van Albada 1996), the
Mileura Widefield Array
(MWA\footnote{http://www.haystack.mit.edu/ast/arrays/mwa/})
and the 21 cm Array (21CMA) (Pen et al. 2004), are ongoing or being planned.

A major challenge to detect the primordial 21 cm signals is from the
contamination of radio signals of the foreground. The brightness
temperature fluctuations of the 21 cm signals would be seriously
contaminated by high-redshift radio point sources, free-free
emissions of the IGM, and the noises from the artificial radio
interference in the VHF band (Di Matteo et al. 2002, 2004; Oh \&
Mack 2003). The power spectrum of the brightness temperature
fluctuations of the redshifted 21 cm map may even be dominated by
the fluctuations of the foreground (Di Matteo et al. 2004). Thus, we
should search for useful features of the primordial 21 signals to
distinguish it from the noisy foreground. For instance, 
the patchy structures of the 21 cm brightness temperature field are 
found to be highly non-Gaussian,  which would be helpful to identify
 the 21 cm signals from the noisy foreground (He et al. 2004, Mellema et al.
2006). Moreover, the correlation between the 21 cm signals and other
emissions of the sources would also be useful for identifying high
redshift sources.

In this context, some studies focused on the 21 cm signals from
individual UV ionizing sources in the reionization epoch. Their
features may provide a direct identification of the ionized patches
of the reionization (e.g., Tozzi et al. 2000). Very
recently, the 21 cm emission and absorption around isolated ionizing
sources has been calculated (Chuzhoy et al. 2006; Cen 2006; Chen \&
Miralda-Escude 2006). However, their results, especially the
time-dependence of the profile of the 21 cm emission and absorption,
	still do not quite agree with each other. This problem comes
partially from the fact that they dropped the time derivative term of
the radiative transfer equation, i.e., the speed of light is taken to
be infinite. This approximation would be reasonable if the
retardation effect is negligible. However, it is not the case for
the 21 cm problem, for which we note that the size of the HII regions
and the age of the sources used in these calculations have already
been close to, or even violated, the retardation constraint $r\leq ct$.
The retardation effect, or the effect of finite velocity of light,
is not trivial. The time and space dependencies of the ionized and
heated regions are substantially affected by the retardation effect (White
et al. 2003; Shapiro et al. 2006a; Qiu et al 2007). Therefore, it is
necessary to re-investigate this problem by fully considering the
kinetics of photons in the phase space.

We will use the WENO algorithm for the radiative transfer
calculation, which has been developed by Qiu et al.
(2006, 2007). The WENO algorithm has been shown to have high order
of accuracy and good convergence in capturing discontinuities and
complicated structures (Shu 2003). The WENO algorithm is able to
incorporate the temperature and species equations and to provide stable
and robust solutions. It is also able to calculate
the fractions of neutral hydrogen and helium within the ionized
spheres, which generally are extremely small, but would be important
to understand the possible correlation of the 21 cm and Ly$\alpha$
signals.

In \S 2 we present the ionization and temperature profiles of the
IGM around a UV ionizing source. \S 3 shows the 21 cm emission and
absorption regions, and their dependency on intensities,
frequency-spectrum and life time of the UV photon sources. 
Discussions and conclusions are presented in \S 4. The algorithm is
summarized in the appendix.

\section{Ionization and Heating of a Point Source}

\subsection{The model}

The whole process of reionization is very complex due to the
inhomogeneous and clustering nature of the structure formation (e.g.
Ricotti et al. 2002; Razoumov \& Norman 2002; Ciardi et al. 2003).
Therefore, the configuration of the ionized regions generally would
be non-spherical. However, in order to be comparable with
previous studies (Chuzhoy et al. 2006; Cen 2006; Chen \&
Miralda-Escude 2006), we will consider the ionization and heating of
a uniform IGM around a point source. This simplified model is
reasonable to reveal the features of the formation and evolution of
the 21 cm emission and absorption regions. The radiative transfer
equation of the specific intensity $J(t, \vec{x},\nu)$ is (e.g.
Bernstein 1988)
%eq1
\begin{equation} \label{eqa-rt}
{\partial J\over\partial t} + \frac{\partial}{\partial x^i} \left
(\dot{x}^iJ \right ) = - k_{\nu} J + S_{\nu} + H\left(\nu{\partial
J\over\partial \nu} - 3J \right ),
\end{equation}
where $\nu$ is the photon frequency and $H$ is the Hubble parameter at
time $t$. Supposing the source is at the center $r=|\vec{x}|=0$, we
have $S(t,|\vec{x}|,\nu)=\dot{E}(\nu)\delta(\vec{x})$, where
$\dot{E}(\nu)d\nu$ is the energy of photons emitted from the central
source per unit time within the frequency range from $\nu$ to
$\nu+d\nu$. The source term can equivalently be described by a boundary
condition as
%eq2
\begin{equation}
\lim_{r\rightarrow 0} 4\pi r^2J(t,r,\nu)=\dot{ E}(\nu).
\end{equation}
We will consider two kinds of frequency sepctrum: 
a power law spectrum $\dot{E}(\nu)=\dot{E_0}(\nu_0/\nu)^{\alpha}$ with the
total intensity (energy per unit time) 
$\dot{E}=\int_{\nu_0}^{\infty}\dot{E}(\nu)d\nu = \dot{E}_0
\nu_0/(\alpha-1)$ and a blackbody spectrum. 
The effect of the expansion of the universe on the number
density is negligible if the time scale concerned is a few Myr at
the epoch $1+z \simeq 20$. However, the effect of the redshift of
the photon energy is essential to calculate the flux of Ly$\alpha$ 
photons (Cen 2006). Therefore,
we will keep the $H$ term on the right hand side of eq.(1).

The absorption coefficient in eq.(1) is given by $k_{\nu}(t,r)
=\sigma_{\nu}(t,r)n_{\rm H}(t,r)$ and
%eq3
\begin{equation}
\sigma_{\nu}(t,r) = \sum_i \sigma_{\rm i}(\nu)\frac{n_i(t,r)}{n_{\rm
H}(t,r)},
\end{equation}
where $i$ runs over the components ${\rm HI}$, ${\rm HeI}$ and ${\rm HeII}$.
$\sigma_{i}(\nu)$ and $n_i(t,r)$ are, respectively, the cross
section of ionization and the number density of component $i$. The
number density $n_i$ is calculated by solving the species
equations involving ionization and recombination processes.
We use the on-the-spot approximation to take into account the effect
of the diffusing
photons and treat the coupling of H and He by the radiation of HeI
recombination continuum and bound-bound transitions
following the method in Osterbrock (1989).
The corresponding coefficients for the ionization and recombination processes
are taken from Fukugita \& Kawasaki (1994) and Theuns et al. (1998).

The IGM is heated by the UV photons with a rate in unit erg
cm$^{-3}$ s$^{-1}$ as
%eq4
\begin{equation}
H =\sum_i n_i\int_{\nu_i}^{\infty}d\nu
   J(t,r,\nu)\sigma(\nu)\frac{\nu-\nu_i}{\nu}\eta(f_e,\nu)
\end{equation}
where $ \nu_i$ is the threshold of frequency to ionize  component
$i$. The fraction of photon-electron energy converted to heat
$\eta(f_e,\nu)$ depends on the free electron fraction $f_e$
and the photon frequency $\nu$. We use the fitting formula of 
$\eta(f_e,\nu)$ given by
Shull \& van Steenberg (1985). The corresponding coefficients for
radiative cooling processes are also taken from Theuns et al. (1998).

The algorithm of the numerical solution of the radiative transfer
equation coupled with the temperature and species equations of 
HI, HeI and HeII are described in Qiu et al. (2006, 2007). A
fifth-order finite difference WENO scheme and a third-order TVD
Runge-Kutta time discretization are applied to the spatial and time
derivative, respectively. A semi-implicit method is used
for the temperature and species equations. The details and tests of the
numerical algorithm are described in Qiu et al. (2006, 2007) and we
give a brief description of the algorithm in the appendix.

\subsection{Ionized sphere}

We first calculate the profile of an ionized sphere around a point
source at $1+z=20$. As an example, we consider a source of a power law
spectrum with $\dot{E}= 7.25\times 10^{44}$ erg s$^{-1}$ and a spctral index
$\alpha=2$, which are the typical values of QSOs at the EUV band (e.g.
Zheng et al. 1997). The initial ionization fraction of hydrogen and
helium is set to be $10^{-4}$, which is the remaining ionization
fraction after recombination, and the initial temperature of the
baryonic gas is set to be 10 K. A $\Lambda$CDM cosmology model with
$\Omega_m=0.27$, $\Omega_b=0.044$, and $h_0=0.7$ is adopted. The
profiles of the ionized sphere at time 0.575, 1.15, 1.725, 2.3 and
3 Myr are shown in Figure 1, in which $f_{\rm HI}=n_{\rm HI}/n_{\rm
H}$, $f_{\rm HII}=n_{\rm HII}/n_{\rm H}$, $f_{\rm HeI}= n_{\rm
HeI}/n_{\rm He}$, $f_{\rm HeII}=n_{\rm HeII}/n_{\rm He}$ and
$f_{\rm HeIII}=n_{\rm HeIII}/n_{\rm He}$, where $n_{\rm H}=n_{\rm
HI}+n_{\rm HII}$ and $n_{\rm He}=n_{\rm HeI}+n_{\rm HeII}+n_{\rm
HeIII}$.

Figure 1 shows that although the main ionization zones have steep
eges (I-front), there are still a small fraction of hydrogen and
helium been ionized beyond the I-fronts. This ionization is produced
by high energy photons which can effectively penetrate into the IGM
further than photons near the ionization thresholds. The fraction $f_{\rm
HII}(r)$ has a long tail before it approaches the initial value
$10^{-4}$; the fraction of HeII shows a similar behavior as that of HII.
The I-fronts of HeII and HeIII generally are different from
that of HII, and the decrease of the fraction of HeIII with $r$ is
much faster than that of HII and HeII.

We define the I-front $r_I(t)$ as the location where $f_{\rm
HI}(r_I, t)=10^{-4}$.  The speed of the
I-front $d r_I(t)/dt$ is plotted as the solid line in Figure 2. It
shows that the evolution of the ionized sphere has two phases
separated by a characteristic time $t_c$. In the first phase
$t<t_c$, $dr_I(t)/dt \simeq c$, and in the second phase $t>t_c$, the
speed decreases with $t$ in a power law, $d r_I(t)/dt\propto
t^{-3/4}$. This feature of two phases was first pointed out by White et al.
(2003) with an approximate solution by neglecting recombinations.
This feature is actually not sensitive to the choice of the number $10^{-4}$;
if we take $f_{\rm HI}(r_I, t)=90\%$, the two phases feature still
holds (Qiu et al. 2007).

Within the I-front, $f_{\rm HI}$ is on the order of
$10^{-5}-10^{-6}$ and corresponds to a column density of the order
of 10$^{14}$ cm$^{-2}$ on a physical length scale of the order of
100 kpc. This value of column density is just of the same order as the
weak Ly$\alpha$ absorptions (Bi \& Davidsen 1997). Therefore, the
ionized sphere is not completely Gunn-Peterson transparent, but may
produce Ly$\alpha$ absorption features on the spectra of background
sources. On the other hand, within the ionization front, $f_{\rm
HeII}\simeq 10^{-4}$, and therefore, the Gunn-Peterson optical depth
of HeII Ly$\alpha$ is larger than that of HI. It leads to significant
fluctuations of the ratio between the two optical depths 
(e.g. Shull et al. 2004; Liu et al. 2006).

We also calculate the ionized sphere of sources with intensities
$\dot{E}= 7.25\times 10^{43}$ and $7.25\times 10^{45}$ erg s$^{-1}$
and a power law index $\alpha=2$. Their I-front speeds are also
plotted in Figure 2. It shows that all the three curves of $d
r_I(t)/dt$ have the two phases. Especially, in the second
phase the power law $d r_I(t)/dt\propto t^{-3/4}$ is
approximately scaled with respect to $(\dot{E})^{1/2}$, 
in other words, the characteristic time $t_c\propto (\dot{E})^{1/2}$.
This scaling can not be found with static solutions of the radiative
transfer equation or by assuming the speed of light is infinite
(e.g. Cen, ApJS, 2002; Whalen \& Norman 2006).

\subsection{Kinetic temperature distribution}

Figure 3 shows the corresponding kinetic temperature distribution
for the ionizing source with $\dot{E}= 7.25\times 10^{44}$ erg s$^{-1}$
and a power law index $\alpha=2$ at time 0.575, 1.15,
1.725, 2.3 and 3 Myr.  The kinetic temperature basically keeps
constant around $3\times10^{4}$ K within the ionized sphere, which
is determined by the balance between the photon-ionization heating
and radiative cooling. An interesting feature is that the temperature
at the center ($r \approx 0$) is lower than
the constant temperature. This is because the ionizing photon flux
at the center is very strong, the number of $f_{\rm HI}$ at $r \sim
0$ is extremely low (see Figure 1), and therefore, the heating rate
is low.

Figure 3 shows that the gas is also substantially heated beyond the
I-front. Actually, the energy spectrum of the radiation is
significantly hardened around the I-front, because soft photons are
effective to ionize the IGM, and hard photons are more effective
to heat the IGM (Qiu et al 2007). After the time $t_c$, soft photons are
exhausted within the ionized sphere $r<r_I(t)$; in contrast, hard
photons are still abundant in the region $r>r_I(t)$, as hard photons
can penetrate further beyond the ionized sphere. The region between
the I-front and the front of light $r=ct$ is a pre-heating layer.
The temperature profile is different at different time and more
elongated at a later time. This is because the speed of the I-front is
much less than the speed of light after $t_c$, while the spped of the 
front of heating is about the same as the speed of light.

We introduce a radius $r_{50}(t)$ by the solution of $T(r,t)=50$ K,
which is about the CMB temperature at $1+z=20$. We call the high
temperature region to be $r_I(t) <r<r_{50}(t)$ and the low temperature
region $r_{50}(t) < r \leq ct$. The latter will play the role of
the 21 cm absorption. Since the speed of the radius $r_{50}(t)$ is
also less than $c$, the low temperature region increases quickly
with time. Therefore, the kinetic temperature distribution shows a
long tail in the range $T < 50$ K. We also calculate the kinetic
temperature distribution of sources with intensities $\dot{E}=
7.25\times 10^{43}$ and $7.25\times 10^{45}$ erg s$^{-1}$ and a
power law index $\alpha=2$. The profiles of the temperature
distribution is substantially $\dot{E}$-dependent.

\section{21 cm Emission and Absorption}

\subsection{Spin temperature and 21 cm brightness temperature}

The 21 cm ($\nu_0=$ 1420 MHz) line emission or absorption associated
with the spin-flip transition of the electron in neutral hydrogen is
significant if the number density $n_{\rm HI}$ of neutral hydrogen
is high and the spin temperature $T_s$, which describes the number ratio
between the spin-up and the spin-down states, deviates from the temperature
 of cosmic microwave background (CMB) $T_{\rm cmb}$. The preheating region
discussed in \S 2.3, i.e., the region between the I-front and the front
of light $r=ct$, would satisfy these two conditions. We may call this
region as a 21cm shell.

We do not consider the bulk motion of the baryonic fluid. Thus, when
$T_s$ is much larger than the hydrogen hyperfine energy $T_{*}=0.06$ K,
the optical depth of
the 21 cm absorption $\tau(z)$ is given by (Wild 1952; Field 1959)
%eq5
\begin{eqnarray}
\tau(z)& = & \frac{3hc^3A_{10}n_{\rm H}f_{\rm HI}(z)}{32\pi\nu_0^2
          k_BT_s(z)H(z)} \\ \nonumber
 & \approx & 2.7 \times 10^{-3} f_{\rm HI}(z)
     \left[ \frac{T_{\rm cmb}(z)}{T_s(z)} \right ] (1+z)^{1/2},
\end{eqnarray}
Since $\tau(z) \ll 1$, the observed brightness
temperature excess (or deficit) for the mean gas density of the universe
at the redshifted frequency $\nu=\nu_0/(1+z)$ is
%eq6
\begin{equation}
\delta T_b
  \simeq  8 \times 10^{-3} f_{\rm HI}
     \left[ \frac{T_s- T_{\rm cmb}}{T_s} \right ] (1+z)^{1/2}
     \hspace{3mm} {\rm K}.
\end{equation}

The decoupling of the spin temperature $T_s$ from the CMB temperature
$T_{\rm cmb}$ could be achieved by the collision of neutral hydrogen
and scattering of Ly$\alpha$ photons, which is the
Wouthuysen-Field mechanism (Wouthuysen 1952; Field 1959). The spin
temperature $T_s$ is then given by a weighted average of the CMB
temperature $T_{\rm cmb}$, the kinetic temperature $T_k$, and the
color temperature $T_{\rm c}$ of background photons at Ly$\alpha$
frequency (Field 1958, 1959)
%eq7
\begin{equation}
T_s=\frac{T_{\rm cmb} + y_\alpha T_{\rm c} + y_c T_k}{1 + y_\alpha +
y_c},
\end{equation}
where $y_\alpha$ is the Ly$\alpha$ photon coupling coefficient and
$y_c$ is the collision coupling coefficient.

At redshift $z=20-30$, the collision coupling is only efficient in overdense
regions (e.g. Furlanetto \& Loeb 2002; Shapiro et al. 2006b), and
is negligible for cosmological mean densities. In the low density IGM, the
Wouthuysen-Field mechanism would play the major role to decouple $T_s$
from $T_{\rm cmb}$. The resonant scattering of Ly$\alpha$ photons by neutral
hydrogen leads to that the color temperature of the radiation
spectrum near Ly$\alpha$ frequency approaches the kinetic temperature of
the baryonic gas. Therefore, if there are enough Ly$\alpha$ photons,
we have $T_{s}\simeq T_k$. There are two important ways to
contribute Ly$\alpha$ photons: One is from the continuum spectrum
photons redshifted to Ly$\alpha$ frequency, which is governed by the
$H$ term in eq.(1). The other one is from the excitation of neutral
hydrogen by high energy electrons, and
we take the fitting formula of the fraction of excitation for high
energy electrons from Shull \& van Steenberg (1985).

\subsection{Evolution of the 21 cm shell}

Figure 4 shows the profiles of the spin temperature $T_s$ and
the brightness temperature $\delta T_b$ for a point source with
$\dot{E}= 7.25\times 10^{44}$ erg s$^{-1}$ and a power law index
$\alpha=2$ at time 0.575, 1.15, 1.725, 2.3 and 3 Myr. The spin
temperature $T_s$ for $T_k>10^4$ K is set to be $T_k$; this will not
affect our results since HI is highly ionized for $T_k>10^4$ K. In
the 21 cm shell, i.e., the region between the I-front and the front
of light $r=ct$, the brightness temperature $\delta T_b\neq 0$. In
the early stage, when the speed of  the I-front $d r_{I}(t)/dt$
is close to the speed of light, there is no 21 cm
shell. The 21 emission and absorption regions become broad when the
difference between $r_I(t)$ and $r=ct$ is significant.

Figure 4 shows that the 21 cm layer always contains an emission region
$\delta T_b > 0$ and an absorption region $\delta T_b < 0$. The
maximum brightness temperature in the emission region is
$\simeq 35$ mK, while the minimum temperature of the absorption
region is $\simeq -100 - -150 $ mK. The absorption region grows
faster than the emission region. Before 1 Myr, the emission region is
comparable with the absorption region, but at 3 Myr, the emission region
is about 2 comoving Mpc while the absorption region exceeds 10
comoving Mpc.

A prominent feature of the brightness temperature in Figure 4
is that the absorption region has a sharp cut-off at the front of light
$r=ct$. This comes from the fact that in our model the
Wouthuysen-Field coupling is efficient within the entire range of the
photon propagation; consequently, in the region $r_{50}(t)< r<ct$,
the Ly$\alpha$ photons are abundant enough to lock $T_s$ with $T_k$.
These photons come from the redshifted photons originally in the
energy range between ionization energy and Ly$\alpha$ line. In this
case, the edge of the 21 cm absorption region is an indicator of the
lifetime of the ionzing source.

\subsection{Dependence on the source intensity}

We also calculate  the profiles of the spin temperature $T_s$ and
the brightness temperature $\delta T_b$ of sources with intensities
$\dot{E}= 7.25\times 10^{43}$ and $7.25\times 10^{45}$ erg s$^{-1}$
and a power law index $\alpha=2$. The results are presented in
Figure 5. Comparing Figures 4 and 5, one can see that the maximum
(emission) and minimum (absorption) of $\delta T_b$ basically are
independent of the source intensity, while the profile of the 21 cm
brightness temperature and its evolution depend strongly on the
intensity.  This is because the heating processes are sensitive to the
evolution of photons in the phase-space, which also cannot be described
by a static approach.

Figure 6 shows the ratio between the sizes of emission and
absorption regions, which decreases with time, because the growth of
the absorption region (brightness temperature $\delta T_b < 0$) is
faster than that of the emission region.  For a source with $\dot{E} =
7.25\times10^{45}$ erg s$^{-1}$, the size of the emission region is
larger than that of the absorption region at an early stage $t< 1$ Myr, while for
sources with $\dot{E} = 7.25\times10^{44}$ and $7.25\times10^{43}$ erg
s$^{-1}$, the size of the emission region is always smaller than that of the
absorption region.

\subsection{Dependence on the photon energy-spectrum}

Figure 7 shows the profiles of the spin temperature $T_s$ and
the brightness temperature $\delta T_b$ for an ionizing source at $1+z=20$
with the spectral index $\alpha=3$ and 4. The intensities for both cases
are normalized to be $\dot{E}= 7.25\times 10^{44}$ erg s$^{-1}$,
i.e., the same as the source of  \S 2.2. Comparing Figures 3 and 7,
we see that the size of the 21 cm emission region of both $\alpha=3$
and 4 is narrower than that of $\alpha=2$. This is due to the lack
of high energy photons for heating. On the other hand, the
absorption regions are more notable than that in the case of $\alpha=2$, as
the sources of $\alpha=3$ and 4 have stronger Ly$\alpha$ photon
fluxes than $\alpha=2$.

The lights from first galaxies may have a thermal spectrum. Let
us consider first galaxies consisting of stars with a blackbody spectrum and
an effective temperature $T\simeq 10^5$ K, which corresponds to
population III stars with mass 200-500 M$_{\odot}$ (Schaerer et al.
2002). Thermal radiation at temperature $T\simeq 10^5$ K does not
contain many photons with energy larger than 100 ev. Therefore, the
pre-heating beyond the I-front is very weak, and the 21 cm region is
small. Figure 8 shows the spin temperature and the brightness
temperature for a first galaxy with the same luminosity as the
source for Figure 4. We see that the size of the 21 cm emission
region of Figure 8 is much narrower than that of Figure 4. However,
the 21 cm absorption region is still significant, because the
Ly$\alpha$ photons are still efficient to support the Wouthuysen-Field
mechanism.

\subsection{Sources of short lifetime}

We now consider a source which is the same as that used for Figures
1 and 3, but the source stops to emit photons after 3 Myr. Figure 9
shows the kinetic temperature $T_k$, the spin temperature $T_s$ and
the brightness temperature $\delta T_b$ at time 3, 3.45, 4.03 and 4.60
Myr. We see that the I-front and $r_{50}(t)$ still move ahead in the
period of 3 to 3.45 Myr, but they slow down after 3.45 Myr. It is
simply due to the fact that there are no more photons to support ionization
and heating. Consequently, the 21 cm emission shell shrinks and almost
disappears at 4.03 Myr.

An interesting result is that the 21 absorption region is alive even
when $t=4.6$ Myr, i.e., 1.6 Myr after the photon source has died out. This is
once again due to the abundance of Ly$\alpha$ photons. When $t>4.03$
Myr, no high energy photons can reach $r>6 $ Mpc, while the
Wouthuysen-Field mechanism is working to the size $r=ct$.
Figure 9 shows that the absorption region at $t=4.60$ Myr is limited
within two sharp cut-offs: $r_1 <r<r_2$. It is easy to check that
$r_1=c\times (4.6 -3) ({\rm Myr})$ and $r_2=c\times (4.60) ({\rm
Myr})$, which is just the range given by the retardation of the
first and last light signals from the source with a luminous lifetime of
3 Myr.

\subsection{Thermal Broadening}

While including the thermal broadening, eq.(5) should be replaced by
%eq8
\begin{equation}
\tau(z) = \frac{3hc^3A_{10}}{32\pi\nu_0^2 k_B}
\int_0^{z_r}\frac{n_{\rm HI}(z')}{T_s(z')H(z')}F(z, z')dz',
\end{equation}
where the factor $F(z, z')$ is the normalized Doppler broadening line
profile as
%eq9
\begin{equation}
F(z, z') = \frac{1}{\sqrt{\pi}b(1+z)}
   e^{-\left (\frac{z'-z}{b(1+z)}\right )^2},
\end{equation}
with $b= (2k_BT/mc^2)^{1/2}$. For gas with temperature $\sim 10^3$
K, we have $b \simeq 1 \times 10^{-5}$, and the comoving scale is
thus $D=1\times 10^{-5}(1+z)c/H(z)$. In the $\Lambda$CDM model at
$(1+z)=20$, we have $D\simeq 0.02$ h$^{-1}$ Mpc. If the thickness of
the 21 cm emission shell is less than $D$, the Doppler broadening
will smooth out the 21 cm emission. As discussed in previous sections,
the 21 cm emission regions are narrow for weak sources and high spectral
index. Therefore, the 21 cm emission would not be observable for
sources with $\dot{E}$ much less than $10^{43}$  ergs s$^{-1}$ or
$\dot{E}\simeq 10^{43-45}$ ergs s$^{-1}$, but $\alpha >2$.

\section{Discussions and Conclusions}

We have studied the formation and evolution of the ionized and heated
region around a UV ionizing source in the reionization epoch. In the
first stage, the ionizing and heating fronts of the IGM are
coincident and propagate with a speed close to the speed of light. The
evolution enters the second stage when the frequency spectrum of UV
photons is hardened due to the loss of photons by reionization. In
this stage, the propagating speed of the I-front is less than
the speed of light, but high energy photons and Ly$\alpha$ photons can
still reach the front of light $r=ct$. The spherical shell
between the I-front $r_I(t)$ and $r=ct$ is the region in which
21 cm signals are produced. The inner shell 
from $r_I(t)$ to $r_{50}(t)$ gives 21
cm emissions; the outer shell from $r_{50}(t)$ to $r=ct$
produces 21 cm absorptions.

Both the 21 cm emission and absorption regions are sensitively
dependent on the intensity, frequency-spectrum and life-time of the
ionizing source. If the intensity is much less than
$\dot{E}\simeq10^{43}$ ergs s$^{-1}$ and the spectral index $\alpha
$ is larger than 3, the emission region is very narrow, and the signal
will probably be smoothed out by thermal broadening. For sources
with intensity $\dot{E} \simeq 10^{44-46}$ ergs s$^{-1}$ and
a spectral index $\alpha =2$, the comoving size $r_{50}(t)-r_I(t)$ is
about 1 - 5 Mpc. The brightness temperature excess upon the CMB is
about 30 mK. Therefore, it yields a radio signal near the band of 70
MHz and is resolvable with observations of angular resolution better
than 0.5 arcmin and spectral resolution better than 40 kHz.

The 21 cm absorption region has very different behaviors from the
emission regions. Absorption region mainly depends on the coupling
of the spin temperature and the kinetic temperature, which is efficient if
there are enough Ly$\alpha$ photons. The ionizing source with
a power law spectrum or thermal spectrum at $T
\simeq 10^5$ can always provide enough Ly$\alpha$ photons until the
largest distance $r=ct$. That is, the size of the absorption region
directly measures the age of the source. For a source with lifetime
$t_s$, the absorption region can survive even when $t$ is larger
than $t_s$. In this region, the brightness temperature deficit with
respect to the CMB is of the order of a few ten mK. The comoving
size of the absorption region can be as large as 10 - 20 Mpc.
Moreover, the small fraction of HI and HeII remained in the ionized
sphere can lead to absorption features on the spectra of background
sources. Therefore, the correlated 21 cm emission, absorption and
Ly$\alpha$ signals would be helpful to identify ionized patches
in the early universe.

Although all the calculations of this paper are on spherical regions,
many features revealed with the spherical profiles
would also hold for non-spherical profiles. This is because many
features of the formation and evolution of 21 cm signals depend
only on the growth of the ionized and heated regions, but not on the details
of their configuration. For instance, it can be shown that 
the growth of a non-spherical ionized region around
clustered sources also consists of two phases as that shown in
Figure 2. In the first phase, the speed of the I-front is also close
to the speed of light, even though in these cases we cannot use one
ionized radius $r_{I}$ to describe the ionized region. Moreover,
similar to the spherical ionized regions, the characteristic time $t_c$
 is longer for clustered sources with higher
intensity $\dot{E}$. Therefore, the stronger the sources, the later
the formation of the 21 cm regions. This property is regardless of the
geometry of the clustered sources (Qiu et al. 2007, in preparation).

In our calculations, the dynamical behavior of cosmic baryonic gas is
ignored. Such an approximation would be reasonable when the speed
of the sound wave are very different from the speed
 of the formation and evolution of the 21 cm emission regions.
Most features of the 21 cm emissions and absorptions 
should still hold when the hydrodynamics of the
baryonic gas is taken into account.

\acknowledgments

We thank the anonymous referee for useful comments and suggestions.
This work is supported in part by the US NSF under the grants
AST-0506734 and AST-0507340. J.L acknowledges the financial support 
from the International Center for Relativistic Astrophysics. L.-L.F
acknowledges support from the National Science Foundation of China
under the grant 10573036.

\appendix{}

\section{Numerical Algorithm}
\subsection{Approximation to the spatial derivative}

To approximate the spatial derivative in equation (1),
the fifth-order finite difference WENO scheme is used.
Specifically, to calculate $\partial J/\partial r$, the variable $\nu$
is fixed and the approximation is performed along the spatial coordinate
$r$
\begin{equation}
\label{shuadd60}
{\partial \over {\partial r}}J(t^n, r_i, \nu) \approx
\frac{1}{\Delta r} \left( {\hat{h}}_{i+1/2} -
{\hat{h}}_{i-1/2} \right),
\end{equation}
where the numerical flux ${\hat{h}}_{i+1/2}$ on a uniform mesh $r_i$
at time step $n$ is obtained with the procedure given below.
We can use the upwind fluxes without flux splitting in the fifth-order
WENO approximation because the wind direction is fixed (positive).

First, we denote
$$
{h_i} = J(t^n, r_i, \nu), \qquad i=-2, -1, ..., {N_r}+2.
$$
The numerical flux from the regular WENO procedure is obtained by
$$
\hat{h}_{i+1/2} = \omega_1 \hat{h}_{i+1/2}^{(1)}
+ \omega_2 \hat{h}_{i+1/2}^{(2)} + \omega_3 \hat{h}_{i+1/2}^{(3)},
$$
where $\hat{h}_{i+1/2}^{(m)}$ are the three third order fluxes on
three different stencils given by
\begin{eqnarray*}
\hat{h}_{i+1/2}^{(1)} & = &
\frac{1}{3} h_{i-2} - \frac{7}{6} h_{i-1}
               + \frac{11}{6} h_{i}, \\
\hat{h}_{i+1/2}^{(2)} & = &
-\frac{1}{6} h_{i-1} + \frac{5}{6} h_{i}
               + \frac{1}{3} h_{i+1}, \\
\hat{h}_{i+1/2}^{(3)} & = &
\frac{1}{3} h_{i} + \frac{5}{6} h_{i+1}
               - \frac{1}{6} h_{i+2},
\end{eqnarray*}
and the nonlinear weights $\omega_m$ are given by
$$
\omega_m = \frac {\tilde{\omega}_m}
{\sum_{l=1}^3 \tilde{\omega}_l},\qquad
 \tilde{\omega}_l = \frac {\gamma_l}{(\varepsilon + \beta_l)^2} ,
$$
with the linear weights $\gamma_l$ given by
$$
\gamma_1=\frac{1}{10}, \qquad \gamma_2=\frac{3}{5},
\qquad \gamma_3=\frac{3}{10},
$$
and the smoothness indicators $\beta_l$ given by
\begin{eqnarray*}
\beta_1 & = & \frac{13}{12} \left( h_{i-2} - 2 h_{i-1}
                             + h_{i} \right)^2 +
         \frac{1}{4} \left( h_{i-2} - 4 h_{i-1}
                             + 3 h_{i} \right)^2  \\
\beta_2 & = & \frac{13}{12} \left( h_{i-1} - 2 h_{i}
                             + h_{i+1} \right)^2 +
         \frac{1}{4} \left( h_{i-1}
                             -  h_{i+1} \right)^2  \\
\beta_3 & = & \frac{13}{12} \left( h_{i} - 2 h_{i+1}
                             + h_{i+2} \right)^2 +
         \frac{1}{4} \left( 3 h_{i} - 4 h_{i+1}
                             + h_{i+2} \right)^2 .
\end{eqnarray*}
$\varepsilon$ is a parameter to avoid the denominator
to become 0 and is taken as $\varepsilon = 10^{-5}$.

\subsection{Numerical calculation in $\nu$}

The computational domain in $\nu$ is on a non-uniform mesh
$$
\nu_j = 2^{\xi_j}\quad with \quad \xi_j=j \Delta \xi, \quad
\Delta \xi = {\log_2}\nu_{max}/N_\nu,\quad
j=0, ... , N_\nu.
$$
The integration of heating rate eq. (4) and photon-ionization rate
is approximated by a fourth order quadrature formula
\begin{equation}
{\int_{\nu_0}^{\infty}}f(x)dx = \Delta x
{\sum_{j=j_0}^{\infty}}{w_j}f(j \Delta x) + O( \Delta x^4),
\end{equation}
where $\nu_0$ is the threshold frequency of ionization, and the weights
$w_j$ are given by
$$
{w_{j_0}}=\frac{3}{8},  \quad {w_{j_0+1}}=\frac{7}{6}, \quad {w_{j_0+2}}
=\frac{23}{24},\quad {w_{j_0+j}}=1, \qquad {\rm for} \ \  j > 2.
$$

\subsection{Time derivative}

The time derivative $\partial J/\partial t$ is calculated using the
third-order TVD Runge-Kutta discretization:
\begin{eqnarray}
\label{rk1}
u^{(1)} & = & u^n + \Delta t L(u^n, t^n)  \\
\label{rk2}
u^{(2)} & = & \frac 3 4 u^n + \frac 1 4 (u^{(1)} + \Delta t L(u^{(1)})) \\
\label{rk3}
u^{n+1} & = & \frac 13 u^n + \frac 23 (u^{(2)}+\Delta t L(u^{(2)})).
\end{eqnarray}
We adapt the multi-timescale strategy and use a semi-implicit scheme in the evolution of
temperature and species. It greatly releases the time step restriction,
therefore saves computational cost. We refer to Qiu et al. (2007) for
more details of the numerical implementation.

%\newpage

%fig1
\begin{figure}
\centering
\includegraphics[height=7in,width=5in]{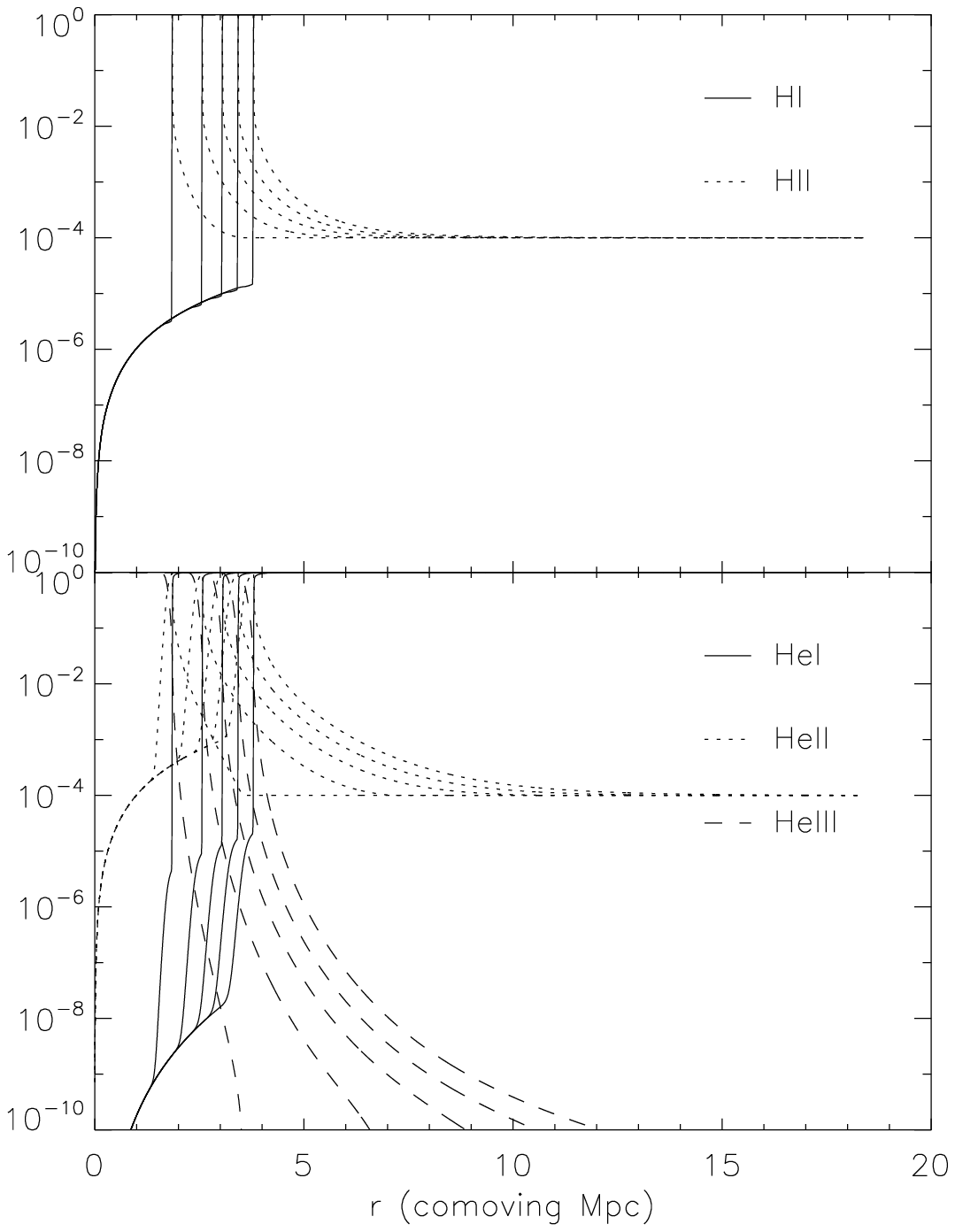}
\caption{$f_{\rm HI}$, $f_{\rm HII}$, $f_{\rm HeI}$, $f_{\rm HeII}$,
and $f_{\rm HeIII}$, the fractions of HI, HII,
HeI, HeII, HeIII around a point source at redshift $1+z=20$ with
an intensity $\dot{E}=7.25\times 10^{44}$ erg s$^{-1}$ and a power low
index $\alpha=2$. From left to right, the time is 0.575, 1.15,
1.725, 2.3 and 3 Myr, respectively.}
\end{figure}

%fig2
\begin{figure}
\centering
\includegraphics{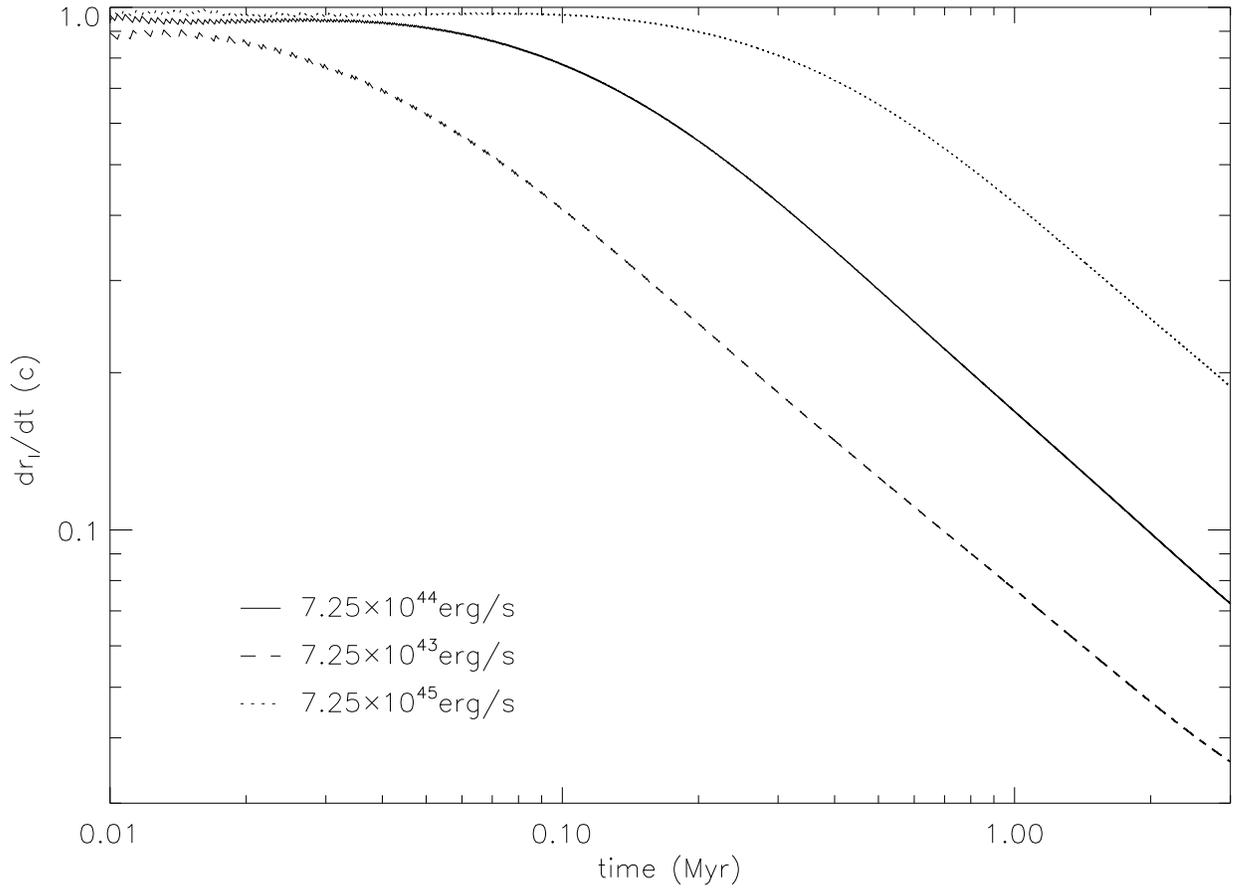}
\caption{The speed of the I-front $dr_I(t)/dt$ for three different
intensities $\dot{E}$. }
\end{figure}

%fig3
\begin{figure}
\centering
\includegraphics{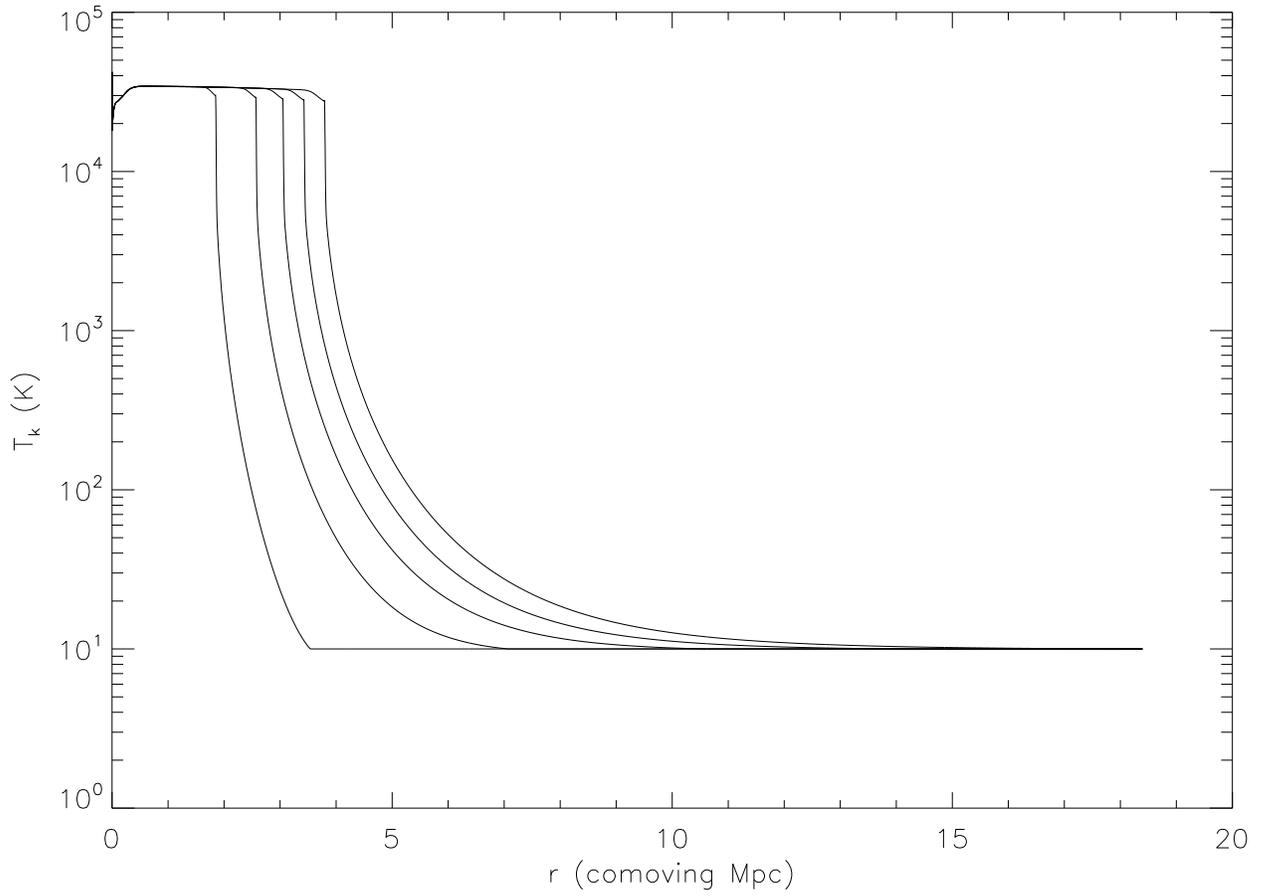}
\caption{The profiles of the kinetic temperature $T(t, r)$ for the same
model as in Figure 1. From left to right, the time is 0.575,
1.15, 1.725, 2.3 and 3 Myr, respectively.}
\end{figure}

%fig4
\begin{figure}
\centering
\includegraphics[height=7in,width=5in]{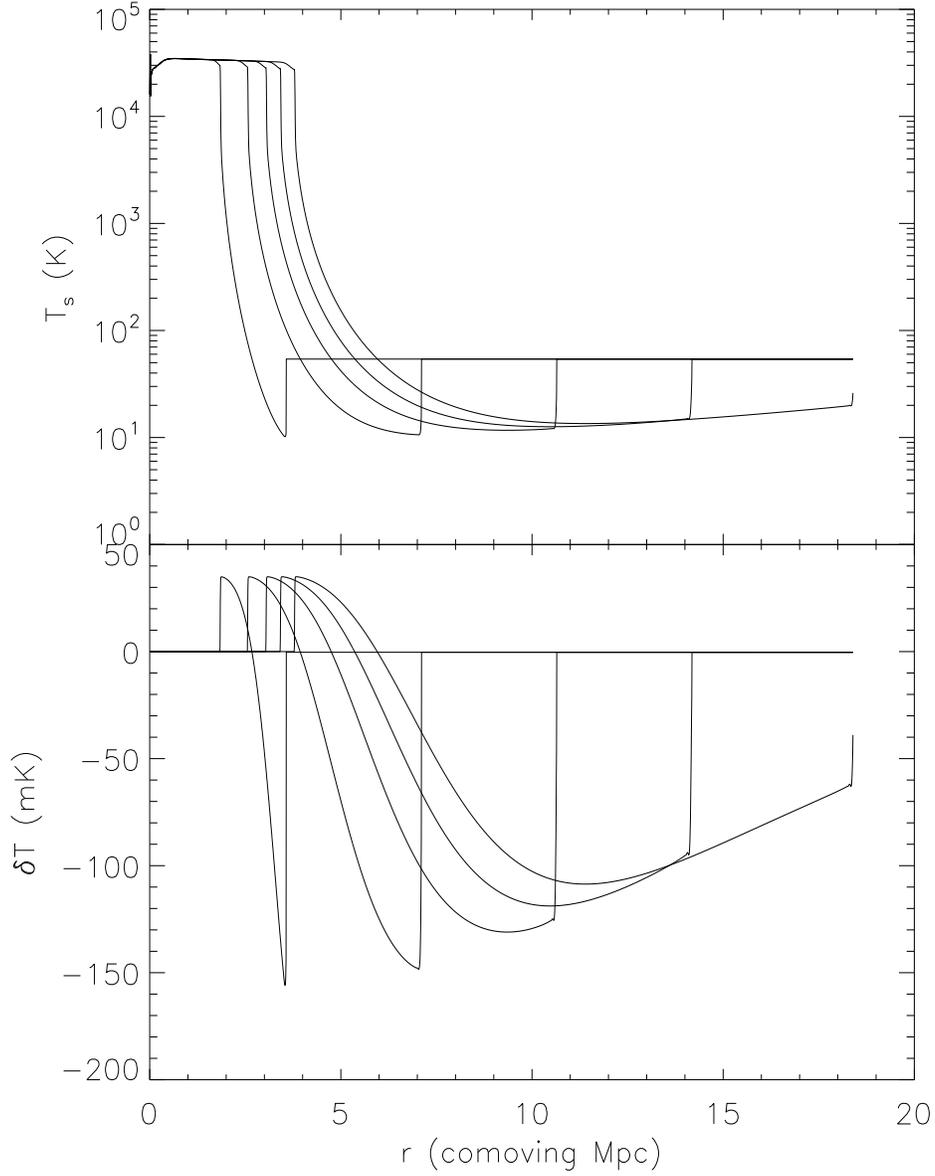}
\caption{The profiles of the spin temperature $T_s$ and the brightness
temperature $\delta T_b$ for the same model as in Figure 1. From left
to right, the time is 0.575, 1.15, 1.725, 2.3 and 3 Myr, respectively. }
\end{figure}

%fig5
\begin{figure}
\centering
\includegraphics[height=5in,width=3.2in]{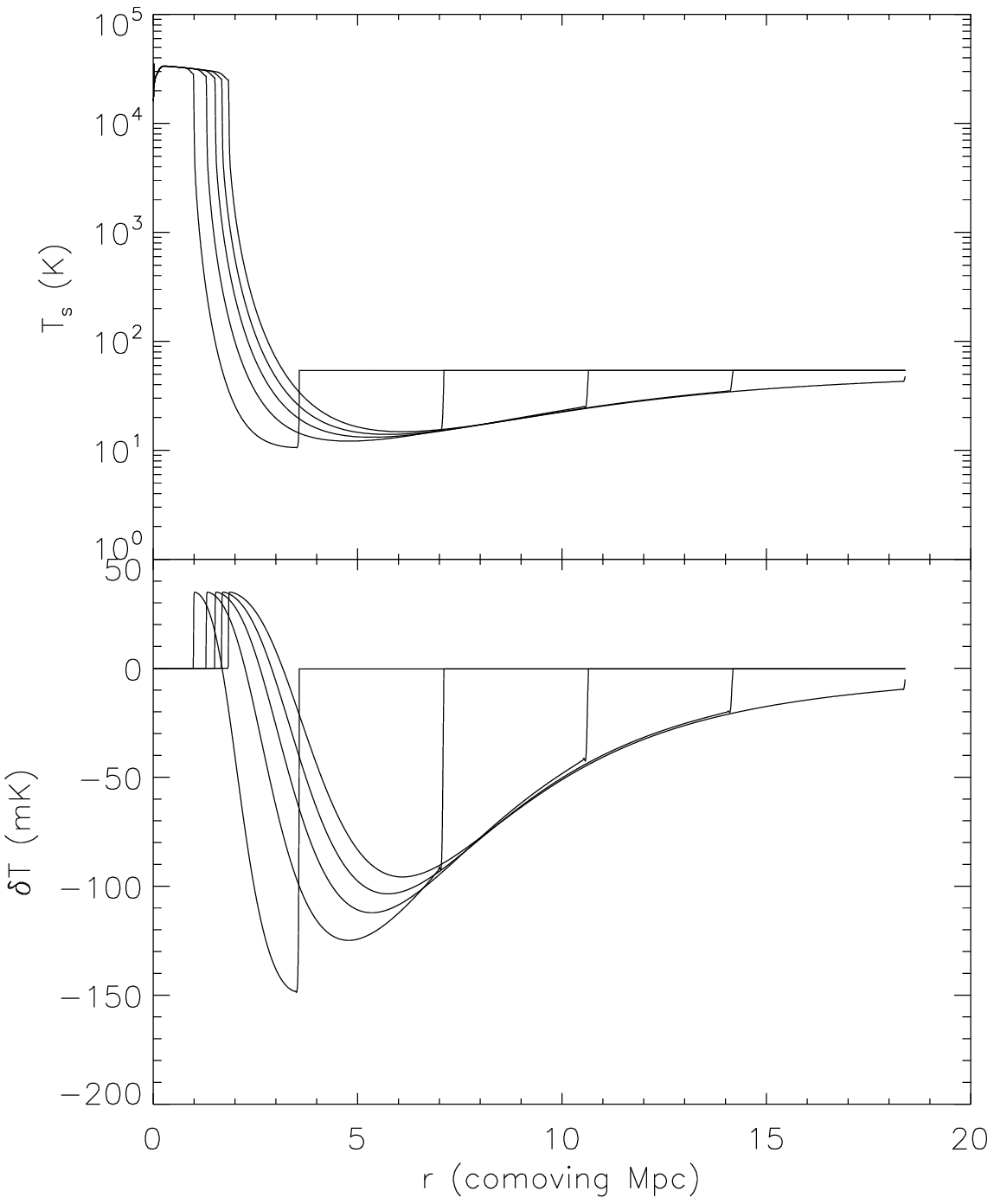}
\includegraphics[height=5in,width=3.2in]{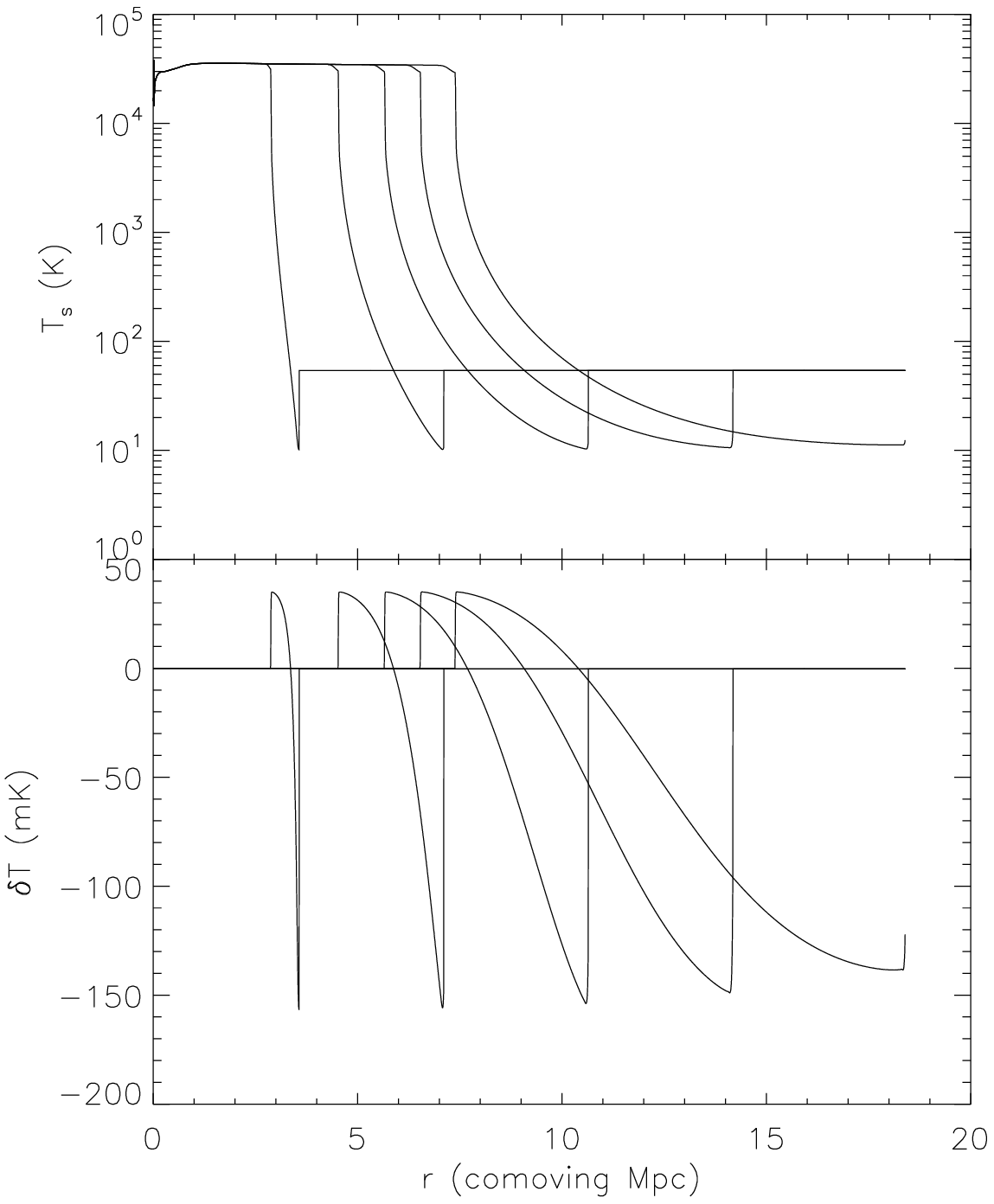}
\caption{The profiles of the spin temperature $T_s$ and the brightness
temperature $\delta T_b$ for point sources with a spectral index
$\alpha=2$, but with intensities $\dot{E}=7.25\times 10^{43}$ (left)
and $7.25\times 10^{45}$  erg s$^{-1}$ (right). In each figure, from
left to right, the time is 0.575, 1.15, 1.725, 2.3 and 3 Myr, respectively.
}
\end{figure}

%fig6
\begin{figure}
\centering
\includegraphics[width=6in]{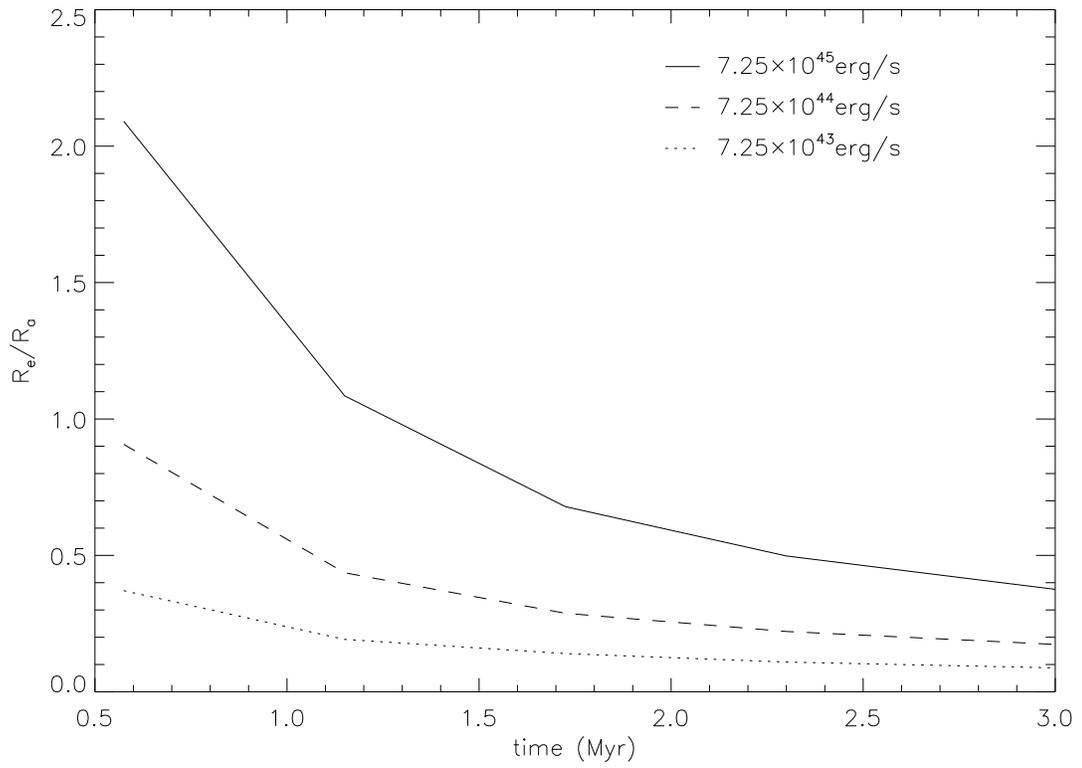}
\caption{The time dependence of the ratio between the size of
emission and absorption regions for three different intensities
    with spectral index $\alpha=2$.
}
\end{figure}

%fig7
\begin{figure}
\centering
\includegraphics[height=5in,width=3.2in]{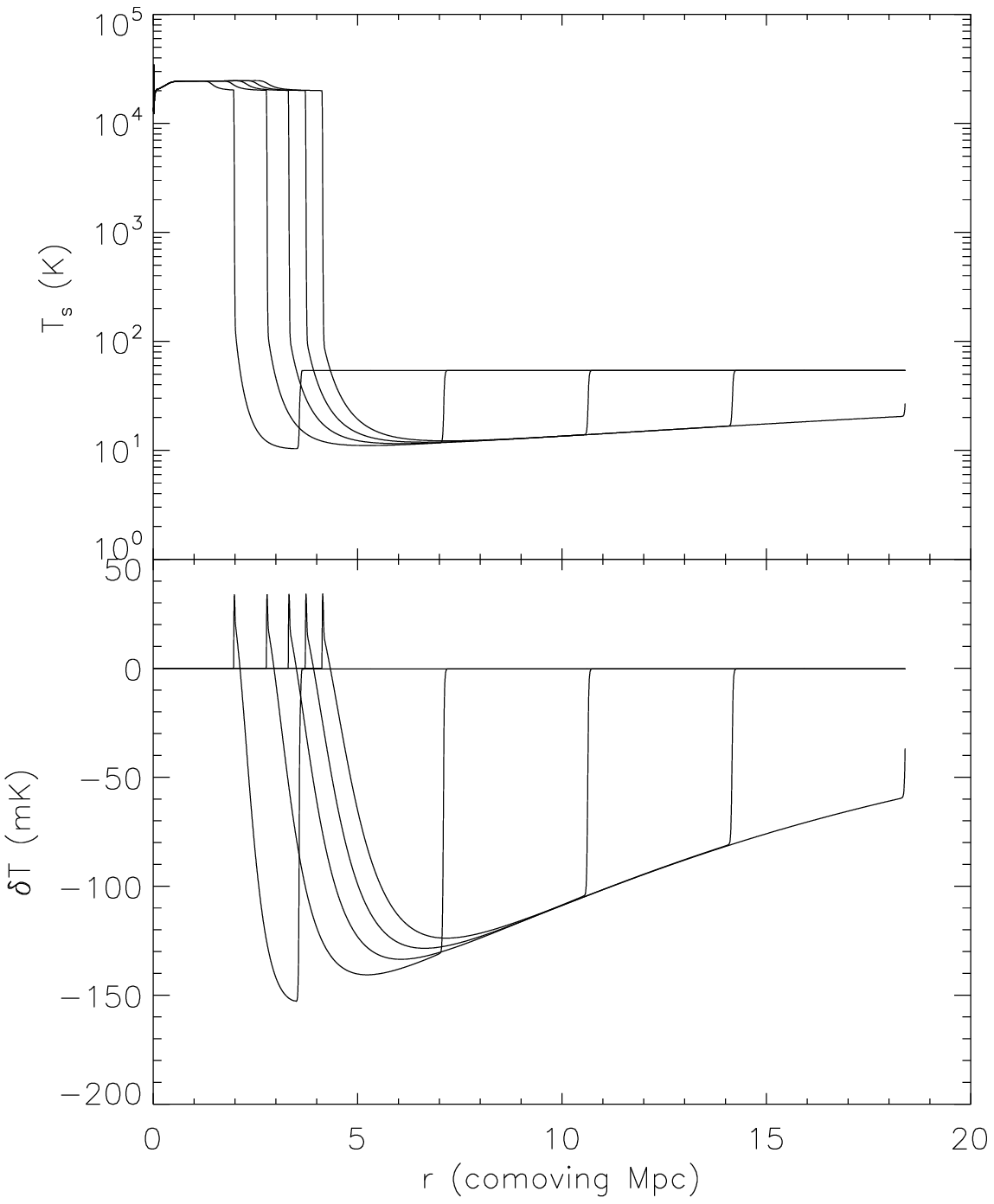}
\includegraphics[height=5in,width=3.2in]{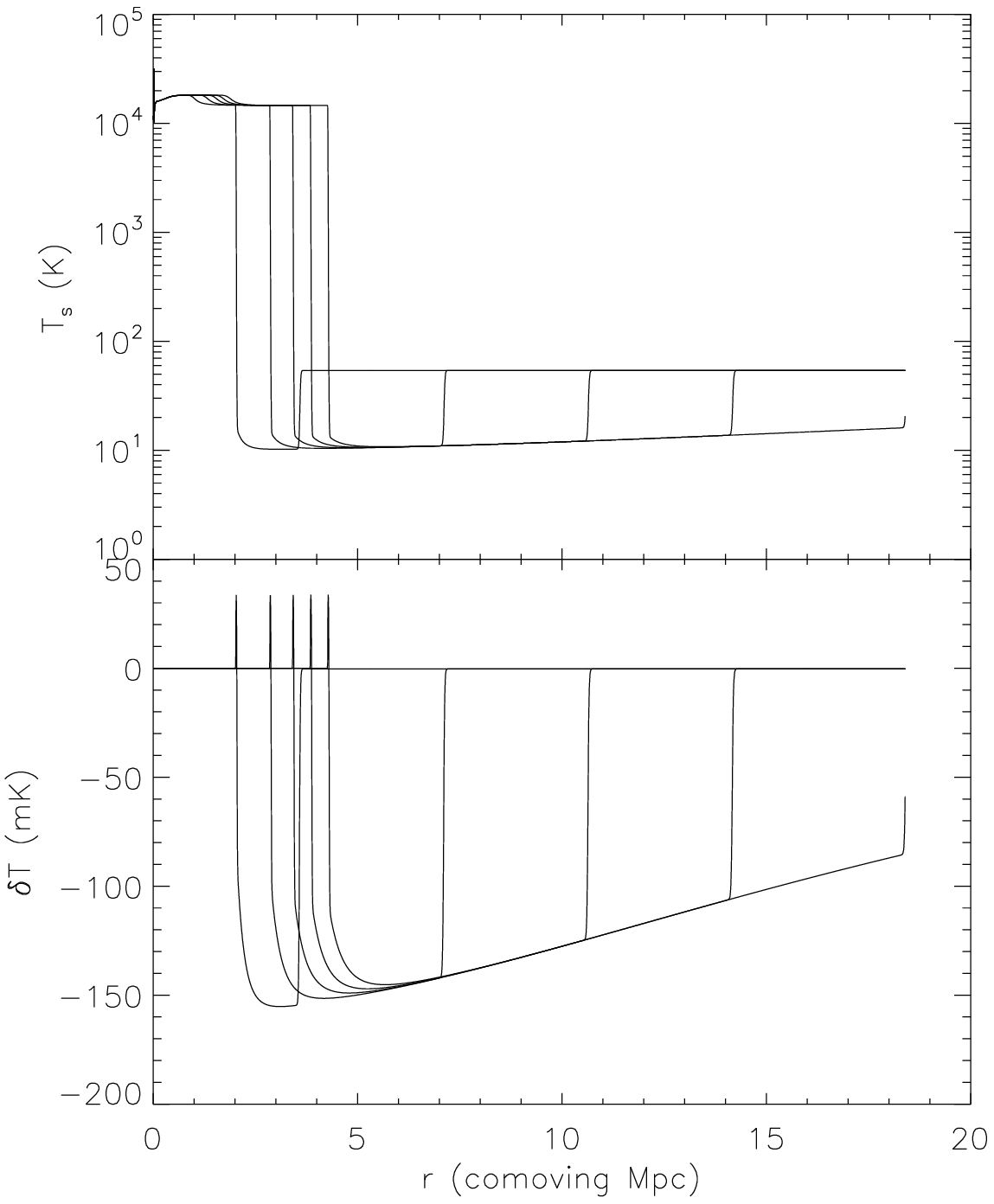}
\caption{The profiles of the spin temperature $T_s$ and the brightness
temperature $\delta T_b$ for a point source with spectral indexes
$\alpha=3$ (left) and 4 (right). The intensities are normalized to
have the same number of ionizing photons as the model in \S 2.2.
In each panel, from
left to right, the time is 0.575, 1.15, 1.725, 2.3 and 3 Myr, respectively.
}
\end{figure}

%fig8
\begin{figure}
\centering
\includegraphics[height=5in,width=3.2in]{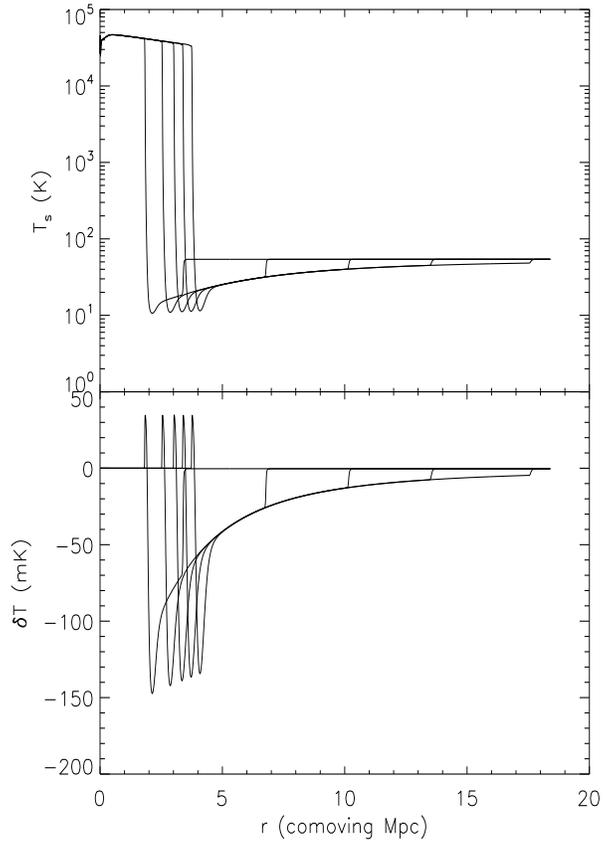}
\caption{The profiles of the spin temperature $T_s$ and the brightness
temperature $\delta T_b$ for a source with a thermal spectrum of
$T=10^5K$ with the same luminosity as in Figure 4. In each panel,
from left to right, the time is 0.575, 1.15, 1.725, 2.3 and 3 Myr, respectively}
\end{figure}

%fig9
\begin{figure}
\centering
\includegraphics[height=5in,width=3.2in]{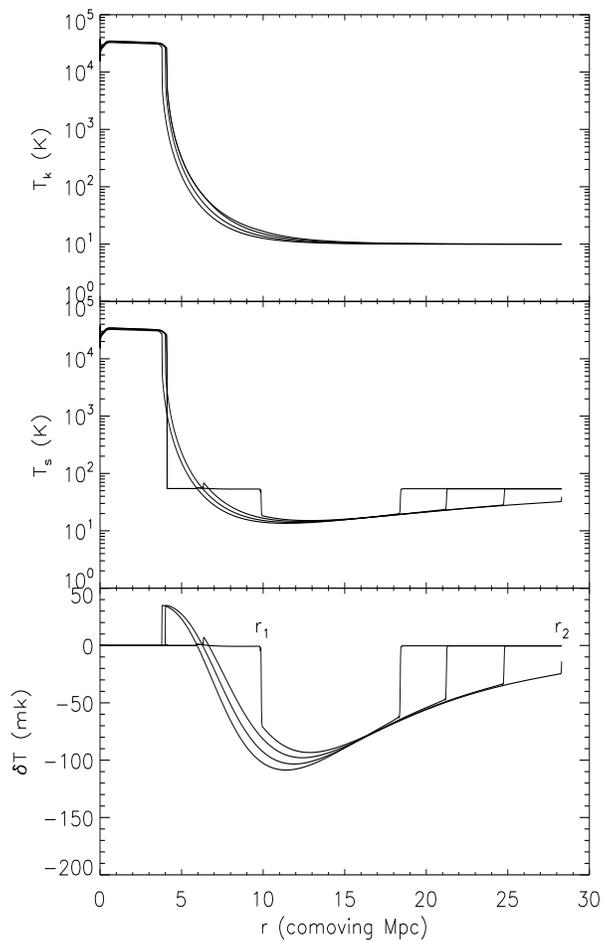}
\caption{The profiles of the kinetic temperature $T_k$,
the spin temperature $T_s$, and the brightness
temperature $\delta T_b$ for the same model
as in Figure 4, but it stops to emit photons after $t=3$ Myr. In each
panel, from left to right, the time is 3, 3.45. 4.03 and 4.60
Myr, respectively. $r_1$ and $r_2$ are the boundaries of the profile $\delta T_b$
at time 4.60 Myr.}
\end{figure}

\end{document}